\documentclass[
    ,final            
  ]
  {aipproc}

\layoutstyle{8x11single}

\usepackage{amssymb,amsmath}
\usepackage{horowitzarXiv}

\begin{document}

\title{Testing pQCD and AdS/CFT Energy Loss at RHIC and LHC}


\author{W.\ A.\ Horowitz}{
  address={Department of Physics, University of Cape Town, Private Bag X3, Rondebosch 7701, South Africa}
}

\begin{abstract}
We compare calculations of jet quenching observables at $\surd s$ = 2.76 ATeV to preliminary LHC data from weak-coupling pQCD and strong-coupling AdS/CFT drag energy loss models. Rigorously constrained to $\surd s$ = 200 AGeV RHIC $\pi^0$ suppression data and introducing no free parameters, the pQCD-based WHDG model simultaneously describes qualitatively the suppression of light hadrons and quantitatively the azimuthal anisotropy (over many centrality classes) of light hadrons and the suppression of $D$ mesons at LHC.  The drag predictions shown here---qualitatively constrained to RHIC data---compare poorly to the measured suppression of $D$ mesons, but the current experimental uncertainties are large.  The double ratio of $D$ to $B$ meson $R_{AA}(p_T)$ should provide a robust experimental measurement to distinguish between the very different assumptions of the strength of interactions in the QGP produced in heavy ion collisions; i.e., whether, from a jet quenching standpoint, the medium is either weakly- or strongly-coupled.
\end{abstract}

\maketitle

\vspace{-.2in}

\section{Introduction}
Since the discovery of a new phase of matter in $\surd s$ = 200 AGeV heavy ion collisions \cite{Gyulassy:2004zy}, the goal of the high energy nuclear physics community is now the quantitative determination of the properties of the quark-gluon plasma (QGP).  For instance, is the medium produced at RHIC and LHC a weakly-coupled gas of the quarks and gluons of the quantum chromodynamics (QCD) Lagrangian or perhaps a strongly coupled fluid made up of emergent collective degrees of freedom, and how does this description change as a function of $\surd s$, $Q^2$, and centrality?  High momentum particles provide the most direct probe of the soft momentum modes present in the QGP created in heavy ion collisions (see \cite{Horowitz:2011gd} and references therein).  Extracting information about the fundamental properties of the QGP requires a comparison between theoretical calculations and experimental data.  Despite its great promise, to date no one energy loss model has provided a consistent picture of the different relevant observables, such as \RAA and \vtwocomma, as a function of \highpt particle mass, $p_T$, centrality, and $\surd s$.  We demonstrate here that calculations based on the WHDG energy loss model \cite{Horowitz:2011gd,Wicks:2005gt}---which is itself based on weak-coupling, perturbative QCD (pQCD) methods---provide a reasonably good, qualitatively consistent description of the preliminary data on the suppression and azimuthal anisotropy of light hadrons and the suppression of $D$ mesons as a function of $p_T$ and centrality at LHC when constrained by a rigorous statistical analysis of central RHIC $\pi^0$ suppression data \cite{Adare:2008cg}.  On the other hand, calculations based on the strong-coupling techniques in an AdS/CFT heavy quark drag model appear to significantly overpredict the central values of the preliminary $D$ meson suppression when qualitatively constrained to the non-photonic electron data from RHIC \cite{Horowitz:2010dm}.  However, the uncertainties related to this heavy meson measurement are quite large, and currently the strong-coupling calculation does not appear to be falsified.  

\vspace{-.2in}
\section{Calculation and Results}
\begin{figure}[!htb]
$\begin{array}{lll}
\includegraphics[width=2.1in]{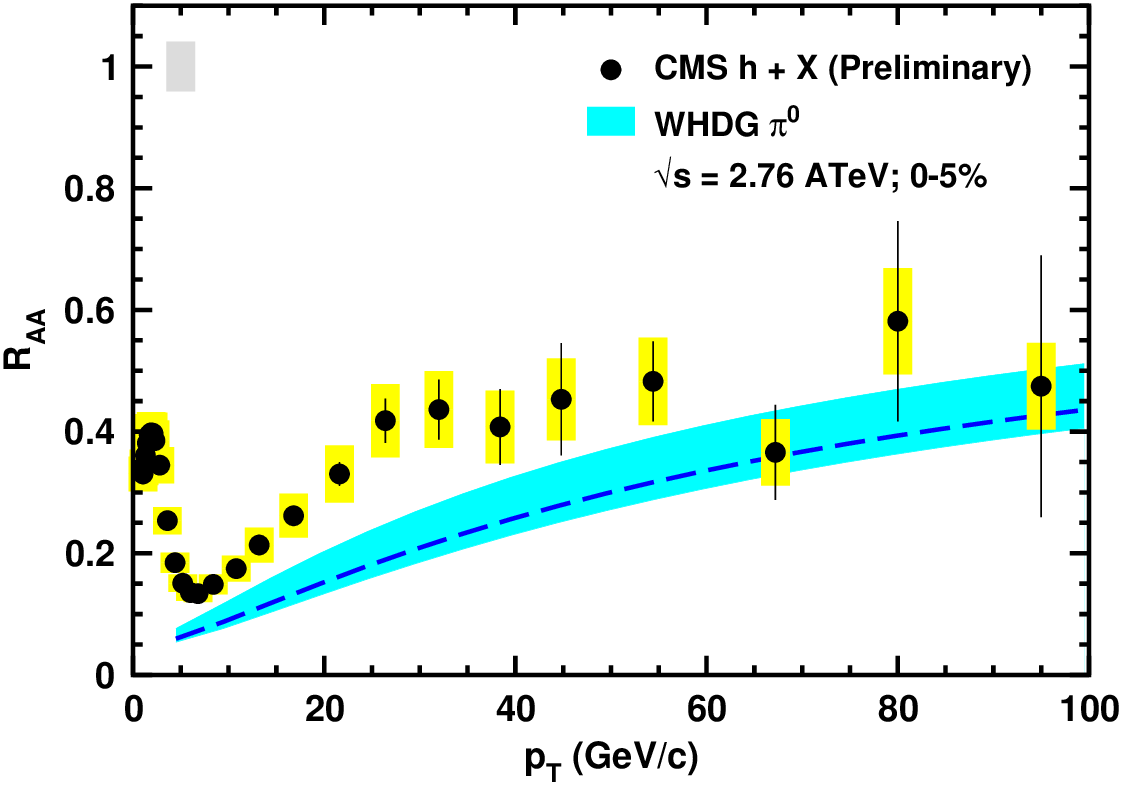} &
\includegraphics[width=2.1in]{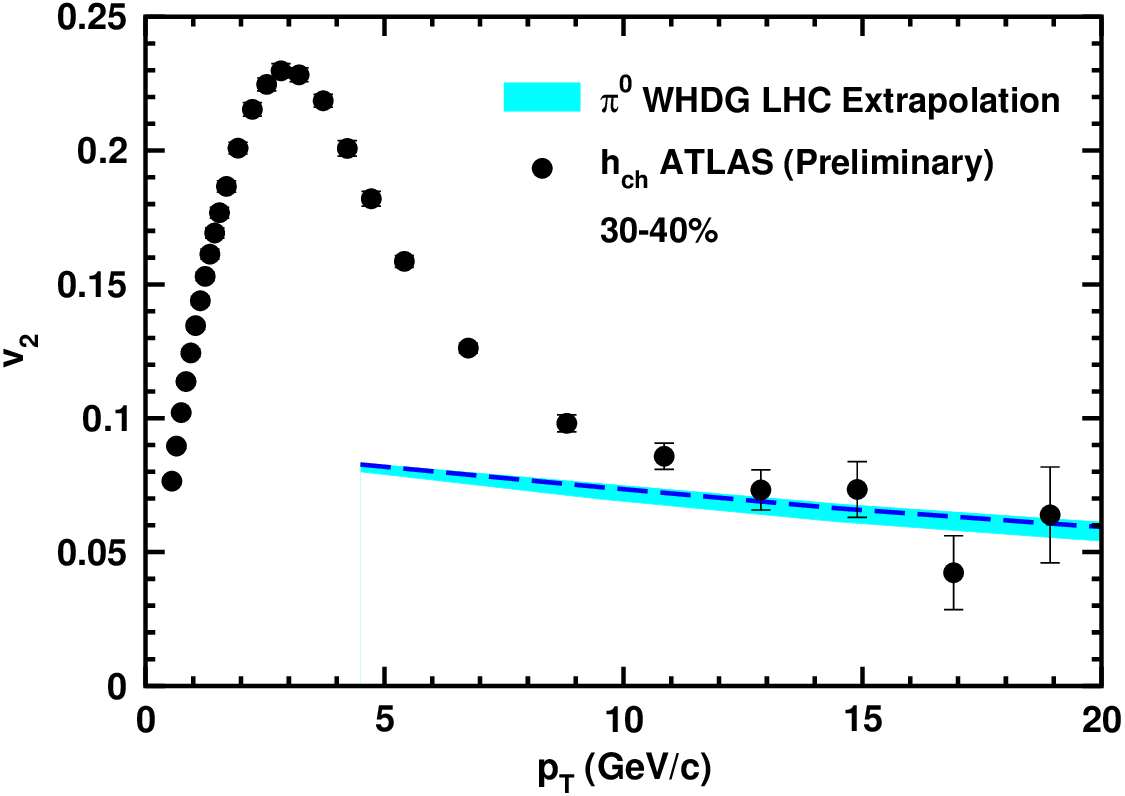} &
\includegraphics[width=2.1in]{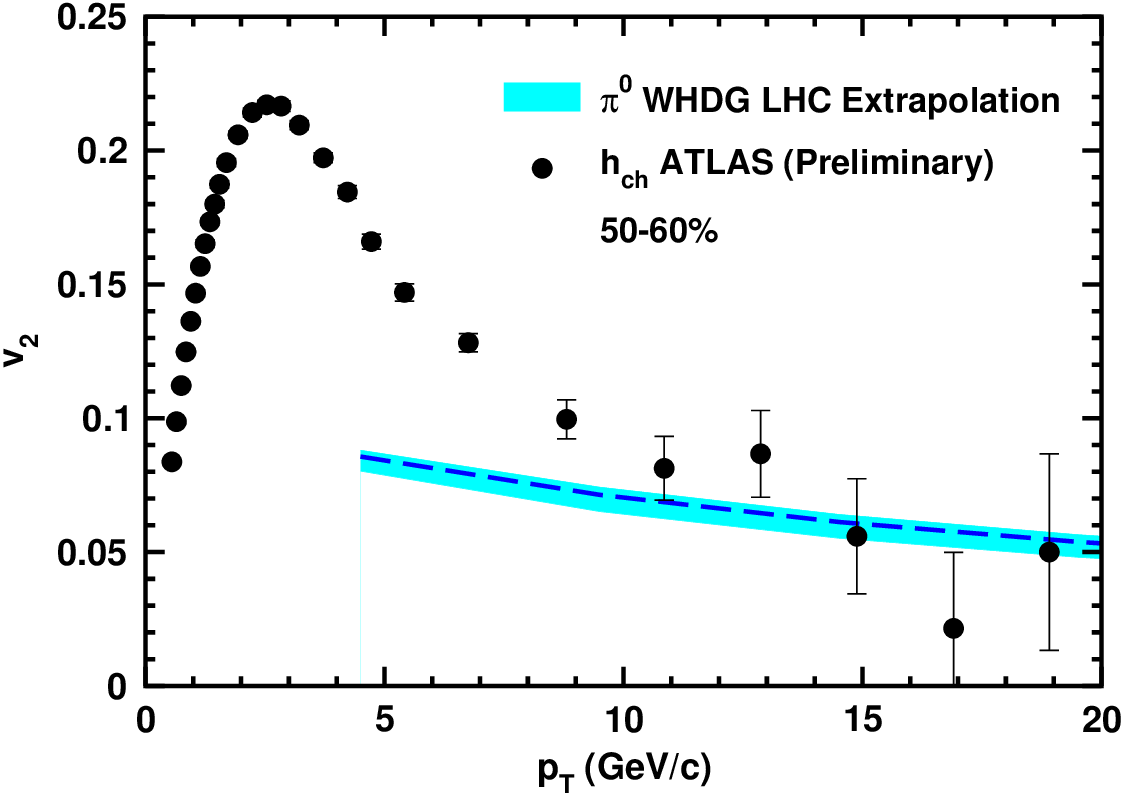}\\[-.1in]
\qquad\mbox{\footnotesize(a)} & \qquad\mbox{\footnotesize(b)} & \qquad\mbox{\footnotesize(c)}
\end{array}$
\vspace{-.5in}
\caption{\label{RAA}(a) Constrained $R^{\pi^0}_{AA}(p_T)$ predictions from the WHDG energy loss model \cite{Horowitz:2011gd,Wicks:2005gt} compared to preliminary CMS $R^{h+X}_{AA}(p_T)$ data \cite{CMSRAA} at $\surd s$ = 2.76 ATeV for 0-5\% central collisions at LHC.  Constrained $v_2^{\pi^0}(p_T)$ predictions from WHDG compared to preliminary ATLAS $v_2^{h^\pm}(p_T)$ data \cite{Jia} at $\surd s$ = 2.76 ATeV for (b) 30-40\% and (c) 50-60\% collisions at LHC. In (a)-(c) the dashed blue line (cyan band) corresponds to the zero-parameter LHC prediction from WHDG for the best-fit constraint (1-$\sigma$ uncertainty on the best-fit constraint) to RHIC data.%
}
\end{figure}

\begin{figure}[!htb]
$\begin{array}{lll}
\includegraphics[width=2.7in]{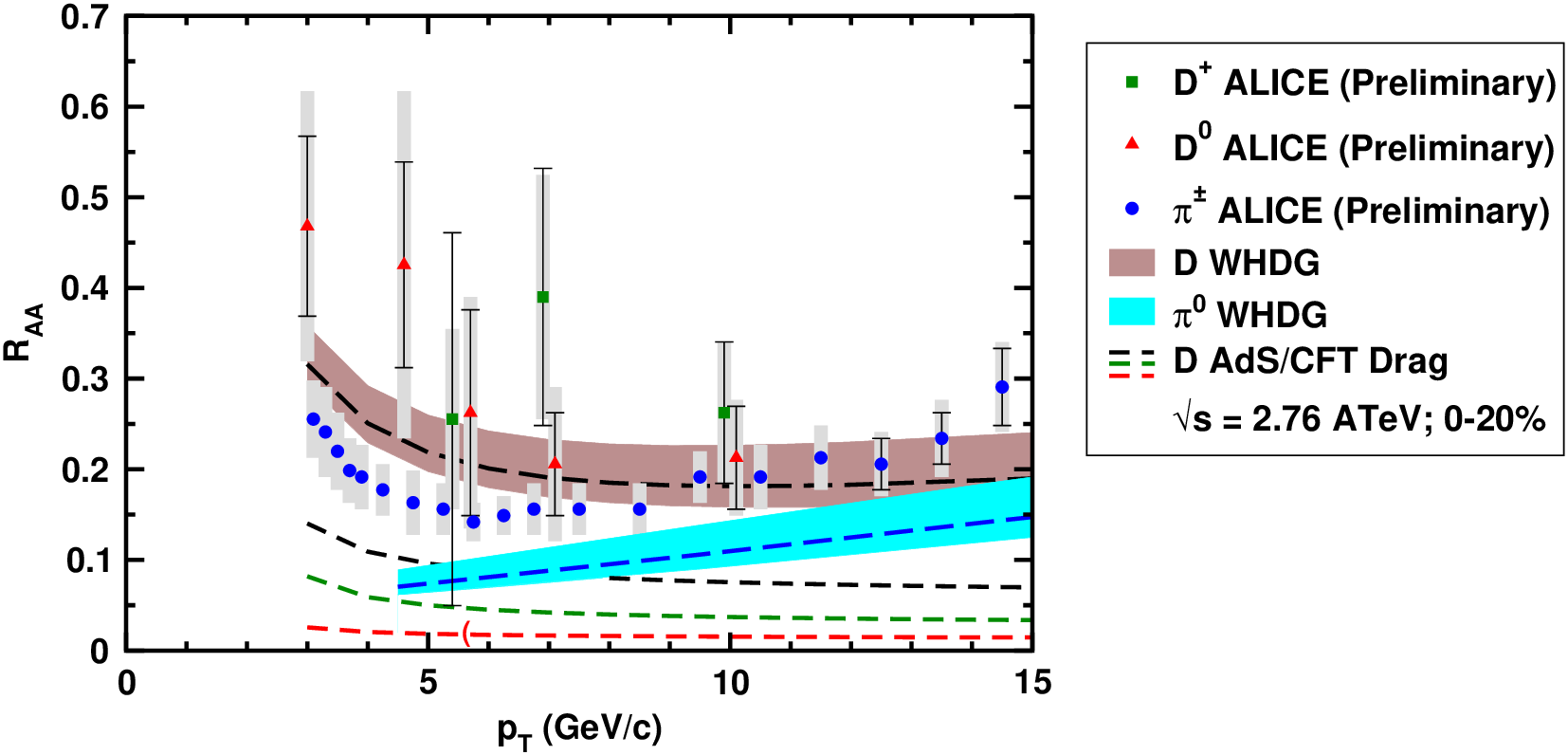} &
\includegraphics[width=1.8in]{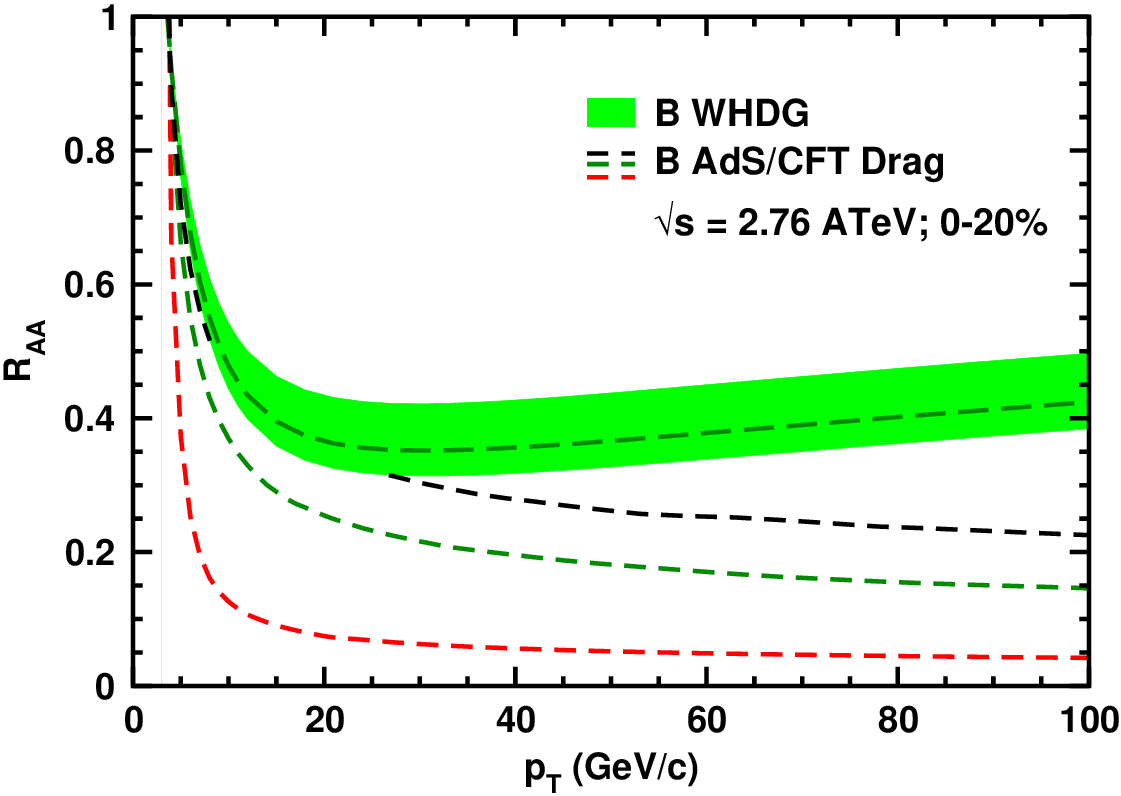} &
\includegraphics[width=1.8in]{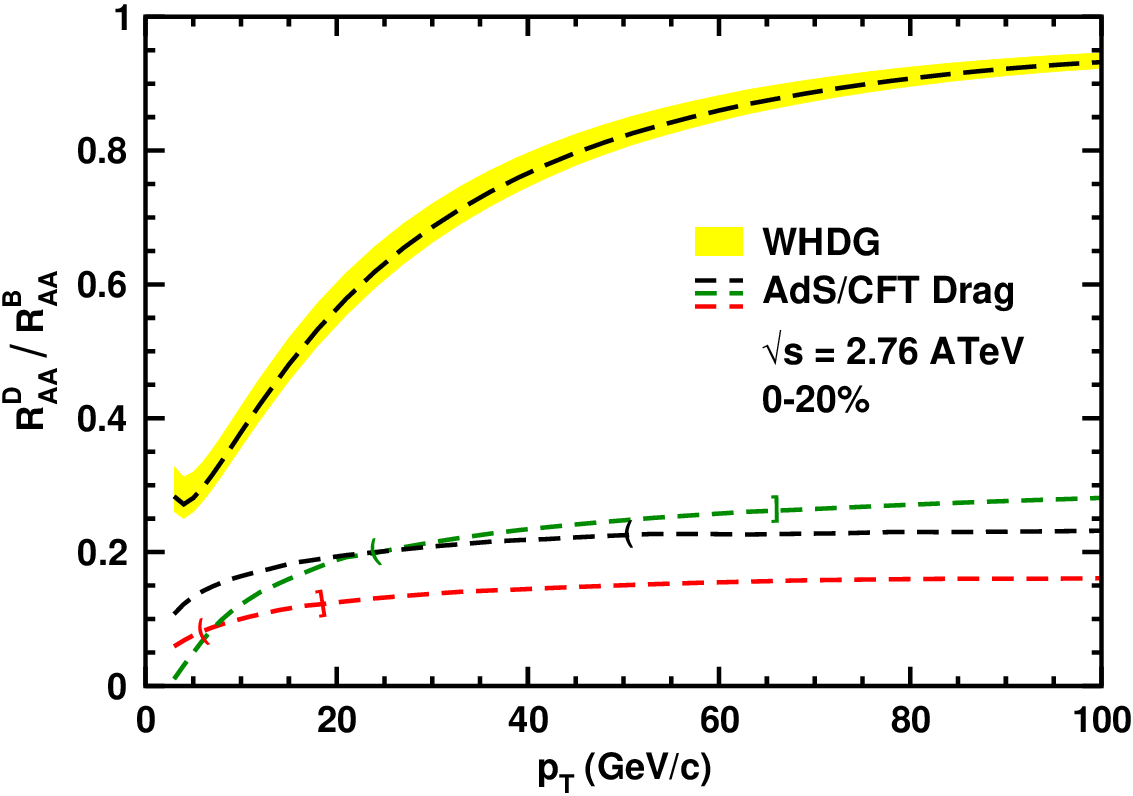} \\[-.1in]
\qquad\mbox{\footnotesize(a)} & \qquad\mbox{\footnotesize(b)} & \qquad\mbox{\footnotesize(c)}
\end{array}$
\caption{\label{RAAheavy}(a) $R^D_{AA}(p_T)$ predictions from an AdS/CFT drag model \cite{Horowitz:2007su} and constrained $R^D_{AA}(p_T)$ and $R^{\pi^0}_{AA}(p_T)$ predictions from WHDG \cite{Wicks:2005gt,Horowitz:2011gd} compared to preliminary ALICE data \cite{Dainese:2011vb} at $\surd s$ = 2.76 ATeV and 0-20\% centrality at LHC.  Constrained (b) $R^B_{AA}(p_T)$ and (c) $R^D_{AA}(p_T)/R^B_{AA}(p_T)$ predictions from WHDG and AdS/CFT drag models at $\surd s$ = 2.76 ATeV and 0-20\% centrality at LHC.%
}
\end{figure}

A generic \highpt energy loss calculation requires a convolution of a hard production spectrum followed by the in-medium energy loss and a fragmentation function.  In WHDG \cite{Horowitz:2011gd,Wicks:2005gt}, the \highpt parton is allowed to lose energy perturbatively through both elastic and inelastic processes in a realistically described, 1D Bjorken-expanding medium.  The radiative energy loss equations were found via the use of standard Feynman diagram techniques \cite{Gyulassy:2000er,Djordjevic:2003zk}, making sure to take into account the crucial fact that the \highpt parton is created at a finite time.  It is important to note that even at top LHC energies, we find elastic energy loss \cite{Thoma:1990fm,Braaten:1991jj,Braaten:1991we} provides a sizable ($\sim$25\%) fraction of the total energy loss \cite{Horowitz:2010dm}.  Using thermal field theory to relate density to temperature and temperature to the Debye mass and mean free path in a QGP, in WHDG the only free parameter becomes the medium gluon density $dN_g/dy$ (with $\alpha_s=0.3$ fixed and geometry set by the colliding nuclei).  As in \cite{Horowitz:2011gd}, in this work we assume that the medium density scales precisely with the measured charged particle multiplicity, $dN_{ch}/d\eta$.  We use the rigorous statistical comparison \cite{Adare:2008cg} of WHDG predictions as a function of $dN_g/dy$ at $\surd s=200$ AGeV to RHIC data to set the proportionality constant between $dN_g/dy$ and $dN_{ch}/d\eta$.  With the published $dN_{ch}/d\eta$ data from ALICE \cite{Aamodt:2010cz}, and continuing to keep $\alpha_s=0.3$ fixed, we can compute all observables at LHC \emph{using no free parameters}.  A sample of such zero parameter predictions is shown in \fig{RAA} and \fig{RAAheavy}.  The prediction associated with the best fit value of the medium density is denoted by a dashed line; the one-sigma uncertainty associated with that best fit parameter extraction is denoted by a band.  In \fig{RAA} (a) we demonstrate a good qualitative agreement of the normalization and momentum dependence between the preliminary CMS $R_{AA}^{h\rightarrow X}$ data \cite{CMSRAA} and $R_{AA}^{\pi^0}$ predictions from WHDG \cite{Horowitz:2011gd}; quantitatively, the calculations overpredict the suppression by, at worst, somewhat more than a standard deviation.  In \figs{RAA} (b) and (c) we show that the WHDG calculations are in excellent quantitative agreement with the preliminary ATLAS \vtwo measurements \cite{Jia} at several centrality bins.  In \fig{RAAheavy} (a) we again see excellent agreement between the WHDG predictions of $D$ meson suppression and the preliminary ALICE data \cite{Dainese:2011vb} within the stated uncertainties.  We provide for future reference WHDG predictions at $\surd s = 2.71$ ATeV of $B$ meson suppression in \fig{RAAheavy} (b) and also of the double ratio of $R^D_{AA}(\eqnpt)/R^B_{AA}(\eqnpt)$ in (c).  Note the rapid convergence of the pQCD-based predictions to unity in \fig{RAAheavy} (c).

\vspace{.2in}

In the AdS/CFT drag model used here the Feynman diagrammatic approach in 4D is replaced by a classical string derivation in 5D \cite{Herzog:2006gh,Gubser:2006bz}.  It was seen in \cite{Horowitz:2010dm} that these strong coupling techniques yield a model whose suppression of non-photonic electrons at $\surd s=200$ AGeV is in both qualitative and quantitative agreement with the data taken at RHIC.  Again assuming that the medium density scales with observed charged particle multiplicity and keeping all other parameters the same produces the AdS/CFT drag curves shown in \fig{RAAheavy} (a)-(c).  The three AdS/CFT drag curves shown explore part of the possible parameter space associated with linking QCD parameters to those in $\mathcal{N}=4$ SYM \cite{Horowitz:2010dm}.  Clearly the drag predictions quantitatively overpredict the suppression of $D$ mesons compared to the preliminary ALICE data \cite{Dainese:2011vb}; however, due to the large uncertainties in that data, the theoretical calculation cannot currently be claimed to be falsified.  Future distinguishing measurements include very different \ptcomma-dependencies for $B$ meson suppression and for the double ratio of $R_{AA}^D(\eqnpt)/R_{AA}^B(\eqnpt)$.  Both of these quantities increase as a function of momentum for pQCD-based calculations.  On the other hand, AdS/CFT drag-based calculations robustly predict a decrease of $R_{AA}^B(\eqnpt)$ as a function of \pt and a near \ptcomma-independence of the double ratio of $R_{AA}^D(\eqnpt)/R_{AA}^B(\eqnpt)$.  One must note that these drag calculations are performed in $\mathcal{N}=4$ SYM, not QCD, and in the limit that $N_c\rightarrow\infinity$ and large $\lambda$.  Additionally the drag calculations assume that the quark mass is very large compared to the temperature of the plasma, an assumption that may not hold particularly well for the charm quark.  Finally, there is a known upper-bound to the velocity of the heavy quark for which the AdS/CFT derivations are applicable \cite{Gubser:2006nz}.  This upper limit is mass-, $\lambda$-, and temperature-dependent.  We give an indication of where corrections to the lowest-order AdS/CFT calculations may set in (at the highest temperature, $T(\vec{x}=\vec{0},\,\tau_0)$, where $\tau_0$ is the thermalization time) with an ``('' on the relevant curves in \fig{RAAheavy} (a)-(c) and where corrections almost certainly set (at the lowest temperature $T=T_c\sim160$ MeV) in with an ``]''. 

\vspace{-.2in}
\section{Discussion and Conclusions}
At RHIC kinematics, no energy loss calculation provided a consistent picture of the multiple observables reported \cite{Horowitz:2011gd}.  In particular, constraining pQCD-based calculations to $\pi^0$ suppression led to a qualitative underprediction of the $v_2^{\pi^0}(p_T\gtrsim9$ GeV/c$)$ \cite{Jia:2011pi} and a quantitative underprediction of non-photonic electron suppression \cite{Wicks:2005gt}.  However, constrained to RHIC data, we find the pQCD-based WHDG energy loss calculation qualitatively describes the light hadron suppression and quantitatively describes the azimuthal anisotropy and heavy meson suppression at LHC, using \emph{no free parameters}.  Nevertheless, one must be cautious in these comparisons to data as there are a large number of theoretical uncertainties that are not yet taken into account in the WHDG---nor fully in any other pQCD-based energy loss---model: higher order effects in opacity \cite{Wicks:2008ta}, coupling \cite{Horowitz:2010dm}, heavy quark mass (divided by parton energy) \cite{Horowitz:2010dm}, collinearity \cite{Horowitz:2009eb}; initial conditions \cite{Renk:2011gj}; energy loss in confined matter \cite{Domdey:2010id}.  On the other hand, the heavy quark drag calculations appear to oversuppress the $D$ mesons compared to the central values of the data, although the results are not completely inconsistent with the data within the current, large experimental uncertainties.  Similar to the perturbative calculations, there are many theoretical uncertainties in the drag calculations that should give one pause when comparing to experimental data; for instance the derivations were performed in $\mathcal{N}=4$ SYM \cite{Ficnar:2011yj} and higher order corrections from, e.g., loops and quark mass to plasma temperature \cite{Chesler:2008uy} have not yet been taken in to account.  We look forward to the future measurements of $B$ meson suppression and the double ratio of $R_{AA}^D(\eqnpt)/R_{AA}^B(\eqnpt)$ at LHC which will help distinguish between the energy loss mechanics dominant in heavy ion collisions.

\end{document}